**Hospitalisation risk for COVID-19 patients infected with SARS-CoV-2 variant B.1.1.7: cohort analysis**

Tommy Nyberg, *research associate* [a]*
Katherine A. Twohig, *senior epidemiology scientist* [b]
Ross J. Harris, *senior statistician* [c]
Shaun R. Seaman, *senior research associate* [a]
Joe Flannagan, *senior epidemiology scientist* [b]
Hester Allen, *principal epidemiology scientist* [b]
Andre Charlett, *head of department* [c]
Daniela De Angelis, *professor of statistical science for health* [a,c]
Gavin Dabrera, *consultant in public health medicine* [b]
Anne M. Presanis, *senior investigator statistician* [a]

[a] MRC Biostatistics Unit, University of Cambridge, Cambridge, United Kingdom.
[b] COVID-19 National Epidemiology Cell, Public Health England, London, United Kingdom.
[c] National Infection Service, Public Health England, London, United Kingdom.

* Corresponding author:
Address: MRC Biostatistics Unit, University of Cambridge, East Forvie Building, Forvie Site, Robinson Way, Cambridge Biomedical Campus, Cambridge, CB2 0SR, United Kingdom.
Email: tommy.nyberg@mrc-bsu.cam.ac.uk
ORCID: 0000-0002-9436-0626

**Abstract**

**Objective** To evaluate the relationship between coronavirus disease 2019 (COVID-19) diagnosis with SARS-CoV-2 variant B.1.1.7 (also known as Variant of Concern 202012/01) and the risk of hospitalisation compared to diagnosis with wildtype SARS-CoV-2 variants.

**Design** Retrospective cohort, analysed using stratified Cox regression.

**Setting** Community-based SARS-CoV-2 testing in England, individually linked with hospitalisation data.

**Participants** 839,278 laboratory-confirmed COVID-19 patients, of whom 36,233 had been hospitalised within 14 days, tested between 23$^{rd}$ November 2020 and 31$^{st}$ January 2021 and analysed at a laboratory with an available TaqPath assay that enables assessment of S-gene target failure (SGTF). SGTF is a proxy test for the B.1.1.7 variant. Patient data were stratified by age, sex, ethnicity, deprivation, region of residence, and date of positive test.

**Main outcome measures** Hospitalisation between 1 and 14 days after the first positive SARS-CoV-2 test.

**Results** 27,710 of 592,409 SGTF patients (4.7%) and 8,523 of 246,869 non-SGTF patients (3.5%) had been hospitalised within 1-14 days. The stratum-adjusted hazard ratio (HR) of hospitalisation was 1.52 (95% confidence interval [CI] 1.47 to 1.57) for COVID-19 patients infected with SGTF variants, compared to those infected with non-SGTF variants. The effect was modified by age (P<0.001), with HRs of 0.93—1.21 for SGTF compared to non-SGTF patients below age 20 years, 1.29 in those aged 20-29, and 1.45—1.65 in age groups 30 years or older. The estimated adjusted risk of hospitalisation within 14 days was 4.7% (95% CI 4.6% to 4.7%) for SGTF and 3.5% (95% CI 3.4% to 3.5%) for non-SGTF patients.

**Conclusions** The results suggest that the risk of hospitalisation is higher for individuals infected with the B.1.1.7 variant compared to wildtype SARS-CoV-2, likely reflecting a more severe disease. The higher severity may be specific to adults above the age of 30.

**What is already known on this topic**
- The SARS-CoV-2 B.1.1.7 variant was discovered in England in December 2020 and has now become the dominant lineage in the country, owing to a higher transmissibility than wildtype SARS-CoV-2.
- Some evidence suggests that B.1.1.7 is associated with more severe disease, but the studies that have found an association with increased mortality may have been limited by confounding due to increased local hospital burden caused by the increased transmissibility of the variant.
- Hospitalisation as a measurement of disease severity is less likely than mortality to be positively confounded by hospital burden.

**What this study adds**
- Based on linkage of nationwide community SARS-CoV-2 testing data with routine hospital admission records, we found that COVID-19 patients who tested positive for the B.1.1.7 variant had a 1.52-fold hazard of hospitalisation within 1-14 days of the first positive test (95% confidence interval 1.47 to 1.57) compared to COVID-19 patients with wildtype variants.
- The results likely reflect a more severe disease associated with the SARS-CoV-2 B.1.1.7 variant, particularly in individuals aged 30 or above.

## 1. Background

Since its discovery in England in December 2020, the SARS-CoV-2 B.1.1.7 variant has been reported in 114 countries globally.[1] In England, the prevalence of B.1.1.7 rapidly increased and it has now become the predominant SARS-CoV-2 lineage,[2] prompting the re-implementation of social and physical distancing measures to control infection rates. These measures included closures of schools, non-essential retail and hospitality outlets and stay at home orders.[3]

Initial concerns around B.1.1.7 emerged from analyses which determined a higher transmissibility.[2,4,5] On 18th December 2020, the variant was re-designated as a Variant of Concern (VOC-202012/01) and subsequent studies have found B.1.1.7 to be associated with higher mortality than other SARS-CoV-2 variants.[6–10]

The burden of COVID-19 on hospital services is a key measure outlined by the UK government on progress in controlling the pandemic, influencing decisions on how quickly social and physical distancing measures can be removed. As B.1.1.7 is now the predominant SARS-CoV-2 lineage in England, any potential increased likelihood of hospitalisation with this variant will impact the national healthcare burden and decisions on lifting restrictions.

Initial assessments of hospitalisation were based on ecological analyses, looking at distribution of variant cases in comparison to the levels of healthcare demand at different geographies.[5,11,12] More recently, a higher risk of admission to critical care has been reported for community-tested COVID-19 patients.[9] One study, based on whole genome sequencing, has reported on the risk of hospitalisation using individual-level follow-up of COVID-19 patients with B.1.1.7 compared to wildtype SARS-CoV-2.[13] However, that study was limited by a moderate sample size due to operational constraints of sequencing, leading to wide confidence intervals for the risk estimates. Hospitalisations linked to individual variant cases based on routine testing data in England, which provide a larger sample size, have not yet been analysed, leaving a gap in the available evidence.

The B.1.1.7 genetic profile includes a deletion of six nucleotides in the S-gene and has been associated with target failures for this gene in PCR testing using a three-gene assay (ORF1ab, N gene and S-gene). While other mutations can also cause an S-gene target failure (SGTF), more than 90% of sequenced SGTF samples since the week commencing 23 November 2020 were confirmed as matching the B.1.1.7 profile.[2] Therefore, SGTF provides an indicator from routine PCR testing that can be used as a proxy for B.1.1.7 and that is more rapidly and widely available than sequencing results.

The aim of this study was to assess whether there is a causal relationship between infection with the B.1.1.7 variant, compared to infection with wildtype SARS-CoV-2 variants, and the risk of hospitalisation. A secondary aim was to re-estimate the mortality risk for patients with the B.1.1.7 variant compared to wildtype variants, that has been reported in previous analyses of the study dataset.[6,7]

## 2. Methods

### 2.1 Identification of confirmed COVID-19 patients by SGTF status

Most SARS-CoV-2 PCR tests in England are performed through the national mass testing programmes.[14] The "Pillar 1" testing programme includes testing by hospital and public health laboratories on request of a health professional for a clinical indication, some testing for public health investigations, and occupational testing of health professionals. The "Pillar 2" testing programme includes large-scale PCR testing of respiratory specimens for SARS-CoV-2 infection in Lighthouse laboratories, predominantly for community-originated testing.[15] These laboratories may receive specimens from testing nationwide depending on demand, and so individual laboratories do not have a fixed geographical coverage. Confirmed COVID-19 patients with SGTF were identified from results uploaded to the Second Generation Surveillance System (SGSS) from the three Lighthouse laboratories using TaqPath assays (Milton Keynes, Alderley Park and Glasgow Lighthouse Laboratories). The identification of these records relied on cycle threshold (CT) values being reported into SGSS from these three laboratories.

We included COVID-19 patients with a positive PCR test from Pillar 2 between 23[rd] November 2020 and 31[st] January 2021 and whose specimen had been analysed in one of the TaqPath assay Lighthouse laboratories. Tests from Pillar 1 were not analysed at Lighthouse laboratories and hence have not routinely been assessed for SGTF status on a national basis. SGTF patients were defined as those patients who had CT values that met the definition for SGTF (ORF1ab and N-gene targets with CT values ≤30 and no values detected for the S-gene). Non-SGTF patients were defined as those patients who had CT values ≤30 at all targets (ORF1ab, N-gene and S-gene). The inclusion period was chosen because SGTF is unspecific to the B.1.1.7 variant and therefore has low positive predictive value when the prevalence of B.1.1.7 is low. Due to the increasing prevalence of B.1.1.7, the positive predictive value of SGTF analysis has been >90% for samples collected in England since the week commencing 23[rd] November 2020.[2] We stopped the inclusion by 31[st] January 2021, because >95% of the analysed samples had SGTF thereafter.[2]

Laboratory data for all included patients were extracted from SGSS and included information on the potential confounders age, sex, ethnicity, area of residence, and index of multiple deprivation (IMD). The dataset was deduplicated to only include each patient's first positive SARS-CoV-2 test.

### 2.2 Assessment of hospitalisation and death

All patients were linked to the Secondary Uses Service (SUS) [16] dataset and the Emergency Care Data Set (ECDS) [17] to obtain information on hospital admissions, as previously described.[18] SUS is an administrative dataset which includes healthcare and hospitalisation data for completed admissions and treatments submitted to NHS Digital. SUS data is not reported until a hospital admission episode is complete (i.e. transfer, discharge or death); ongoing hospitalisations are not included in this dataset. This information can be complemented with ECDS, a similar administrative dataset recording attendances at Emergency Departments, including hospital admissions following emergency room attendance, thus providing another route to capture hospitalisation earlier than in SUS.

SUS/ECDS data were extracted and linked with the laboratory data, including hospitalisation records up to the 19[th] May 2021. If a COVID-19 patient was detected in SUS, they were classified as hospitalised with COVID-19 if they entered the hospital between 1 and 14 days following their specimen date. If the individual was detected in ECDS only, they were classified as hospitalised if they had a discharge status of "Admitted" or "Transferred" and their attendance date was between 1 and 14 days following their specimen date. The timeframe of 1-14 days was based on a preliminary descriptive analysis ignoring SGTF status, that indicated that most hospitalisations occurred within 14 days; we explored including later hospitalisations in an additional analysis. Data were not consistently available on the reason for hospital admission and we therefore included all hospitalisations recorded within this timeframe. Individuals who first tested positive on or after their hospital admission were excluded to avoid bias of healthcare-acquired SARS-CoV-2 infections or testing at admission for non-COVID-19-related hospitalisation for infection control purposes. Similarly, individuals in hospital within 6 weeks prior to testing positive were excluded from analysis, due to the possibility of hospital-acquired infection.

The data were further linked to a dataset of COVID-19 deaths collated by Public Health England (PHE) from the following streams: (i) deaths occurring in hospitals, notified to NHS England by NHS trusts, (ii) deaths among COVID-19-positive individuals, notified to PHE Health Protection Teams during outbreak management, (iii) laboratory test result reports linked with death reports from NHS records, and (iv) death registrations where COVID-19 was mentioned on the death certificate that could be retrospectively linked to a laboratory confirmed COVID-19 test.[19]

**2.3 Potential confounders**

The risk of COVID-19 hospitalisation in England has been reported to be positively associated with age, male sex, deprivation, and Black or Asian minority ethnicity.[20] A previous analysis noted that the prevalence of SGTF variants among COVID-19 patients in England was higher in younger than older age groups.[7] The B.1.1.7 variant was discovered in South East England and the outbreak was initially localised to this and neighbouring regions;[4,5] hence, the prevalence of the B.1.1.7 variant varied by region and calendar period. We therefore treated age, sex, deprivation, ethnicity, region of residence, and date of specimen as potential confounders due to their known associations with the exposure and/or outcome.

**2.4 Patient and public involvement**

This study was observational and based on data from routine healthcare records. No patients were directly involved in the study.

**2.5 Ethical considerations**

This surveillance was performed as part of Public Health England's responsibility to monitor COVID-19 during the current pandemic. Public Health England has legal permission, provided by Regulation 3 of The Health Service (Control of Patient Information) Regulations 2002 to process confidential patient information (http://www.legislation.gov.uk/uksi/2002/1438/regulation/3/made) under Sections 3(i) (a) to (c), 3(i)(d) (i) and (ii) and 3(3) as part of its outbreak response activities. As such this work falls outside the remit for ethical review.

## 2.6 Statistical analysis

*2.6.1 Hospitalisation*

The primary analysis was a stratified cohort analysis with the aim to estimate the hazard ratio (HR) of hospitalisation within 1-14 days for patients who tested positive with SGTF compared to non-SGTF variants, while adjusting for confounding. For this outcome, we followed the patients from the date of their first positive test until the date of hospitalisation if within 14 days, or censored them at the date of death or 14 days after the date of first positive test, whichever occurred first. We estimated age-group-specific absolute risks of hospitalisation for SGTF and non-SGTF patients based on models stratified by age group.

Using Cox regression, we estimated the crude HR of hospitalisation within 1-14 days after testing positive for SARS-CoV-2 for SGTF compared to non-SGTF patients. We then estimated an adjusted HR based on stratification by groups defined by intersecting the potential confounders: 10-year age group, sex, ethnicity, IMD quintile, region of residence (PHEC), and week of specimen. This model (henceforth referred to as the "base model") included SGTF status as a binary covariate, and additionally included strata-specific linear terms for the quantitative covariates age, IMD rank and calendar date to account for residual confounding from these covariates within each stratum.

We tested for deviation from the proportional hazard assumption using Schoenfeld tests, and visually assessed the assumption by examining log-log-transformed Kaplan-Meier plots for SGTF status and each potential confounder. Because stratification may result in loss of observations, we assessed the impact on the HR estimate by omitting each of the potential confounders from the stratification set one by one.

Next, we assessed whether the HR for SGTF was modified by the potential confounders. This was based on likelihood ratio tests between the base model in which the effect of SGTF was assumed constant, and the corresponding models that additionally included interaction terms between SGTF status and each stratification covariate.

In additional analyses, we (1) refitted the base model, stratified by lower tier local authority (LTLA; 316 areas) of residence instead of PHEC (9 regions); and (2) assessed the impact on the results by refitting the base model considering hospitalisations within 1-60 days. For the latter analysis, we allowed for time-variation in the HR for SGTF by fitting a model that assumed piecewise constant HRs by week since positive test, and tested for time-variation using a likelihood ratio test of this model compared to the model with constant HR.

To estimate adjusted absolute risks of hospitalisation within 14 days by SGTF status, we fitted a Cox regression model stratified by age group and SGTF status (henceforth referred to as the "absolute risk model"), including main effects for the remaining potential confounders. This model allows estimation of absolute risks, under the assumption of multiplicative effects of each covariate on the hazard. We estimated age-group-specific absolute risks by SGTF status, evaluated at the mean value of the other potential confounders over all patients. To estimate the corresponding overall absolute risk, we averaged the age-group-specific estimates over all individuals by SGTF status. We used bootstrapping (1000 samples) to estimate 95% CIs for the absolute risks.

*2.6.2 Mortality*

In a secondary analysis, we aimed to estimate the adjusted HR of death within 28 days of a positive test for SGTF compared to non-SGTF patients. For this, the patients were followed from the date of their first positive test until the date of death if deceased within 28 days, or otherwise they were censored after 28 days. Because the outcome was rarer compared to hospitalisation, we estimated the HR based on Cox regression stratified by age, region of residence and week of specimen, and including main effects for the other potential confounders.

We used Stata software (release 14.1, StataCorp LLC, College Station, TX, USA) for the statistical analysis.

## 3. Results

### 3.1 Description of COVID-19 patients by SGTF status

Within the study period, there were 839,278 confirmed SARS-CoV-2 cases reported from TaqPath assay Lighthouse laboratories with valid SGTF status, and who were not hospitalised within 6 weeks prior to testing positive: 592,409 SGTF patients and 246,869 non-SGTF patients. These patients represented 41.0% of all confirmed cases during that time. Table 1 shows the characteristics of the SGTF and non-SGTF patients. The mean age of SGTF patients was 37.6 years, and in the non-SGTF group the mean age was 37.8 years. There were marked differences by region, with higher proportions of SGTF patients in London, East of England and the South East; and differences over time, with the majority of SGTF patients occurring towards the end of December 2020 and the start of 2021, while non-SGTF patients decreased over time.

**Table 1** Characteristics of SGTF and non-SGTF patients.

|  | Overall | SGTF | Non-SGTF |
|---|---|---|---|
|  | Count (%) | Count (%) | Count (%) |
| **Total** | 839,278 | 592,409 | 246,869 |
| **Age** | | | |
| <10 | 44,216 (5.3%) | 31,935 (5.4%) | 12,281 (5.0%) |
| 10-19 | 93,730 (11.2%) | 63,084 (10.6%) | 30,646 (12.4%) |
| 20-29 | 160,857 (19.2%) | 115,296 (19.5%) | 45,561 (18.5%) |
| 30-39 | 165,570 (19.7%) | 118,229 (20.0%) | 47,341 (19.2%) |
| 40-49 | 144,265 (17.2%) | 102,684 (17.3%) | 41,581 (16.8%) |
| 50-59 | 132,211 (15.8%) | 93,468 (15.8%) | 38,743 (15.7%) |
| 60-69 | 63,897 (7.6%) | 44,709 (7.5%) | 19,188 (7.8%) |
| 70-79 | 23,203 (2.8%) | 15,726 (2.7%) | 7,477 (3.0%) |
| 80+ | 11,329 (1.3%) | 7,278 (1.2%) | 4,051 (1.6%) |
| **Sex** | | | |
| Female | 436,049 (52.0%) | 305,230 (51.5%) | 130,819 (53.0%) |
| Male | 403,229 (48.0%) | 287,179 (48.5%) | 116,050 (47.0%) |
| **PHEC Region** | | | |
| East Midlands | 44,407 (5.3%) | 22,913 (3.9%) | 21,494 (8.7%) |
| East of England | 84,454 (10.1%) | 71,250 (12.0%) | 13,204 (5.3%) |
| London | 169,606 (20.2%) | 141,864 (23.9%) | 27,742 (11.2%) |
| North East | 48,227 (5.7%) | 28,120 (4.7%) | 20,107 (8.1%) |
| North West | 151,897 (18.1%) | 94,050 (15.9%) | 57,847 (23.4%) |
| South East | 128,844 (15.4%) | 109,794 (18.5%) | 19,050 (7.7%) |
| South West | 26,382 (3.1%) | 17,235 (2.9%) | 9,147 (3.7%) |
| West Midlands | 117,577 (14.0%) | 74,730 (12.6%) | 42,847 (17.4%) |
| Yorkshire & Humber | 67,884 (8.1%) | 32,453 (5.5%) | 35,431 (14.4%) |
| **Ethnicity** | | | |
| White | 615,523 (73.3%) | 430,930 (72.7%) | 184,593 (74.8%) |
| Asian | 124,156 (14.8%) | 85,829 (14.5%) | 38,327 (15.5%) |
| Black | 36,778 (4.4%) | 28,604 (4.8%) | 8,174 (3.3%) |
| Mixed | 17,880 (2.1%) | 13,330 (2.3%) | 4,550 (1.8%) |
| Other | 31,491 (3.8%) | 23,816 (4.0%) | 7,675 (3.1%) |
| Unknown | 13,450 (1.6%) | 9,900 (1.7%) | 3,550 (1.4%) |
| **Presence of symptoms** | | | |
| No | 121,651 (14.5%) | 88,583 (15.0%) | 33,068 (13.4%) |
| Yes | 717,627 (85.5%) | 503,826 (85.0%) | 213,801 (86.6%) |
| **Index of multiple deprivation** | | | |
| Quintile 1 – most deprived | 202,957 (24.2%) | 132,643 (22.4%) | 70,314 (28.5%) |
| Quintile 2 | 190,807 (22.7%) | 136,868 (23.1%) | 53,939 (21.8%) |
| Quintile 3 | 162,121 (19.3%) | 117,727 (19.9%) | 44,394 (18.0%) |
| Quintile 4 | 148,798 (17.7%) | 106,589 (18.0%) | 42,209 (17.1%) |
| Quintile 5 – least deprived | 134,595 (16.0%) | 98,582 (16.6%) | 36,013 (14.6%) |
| **Specimen date** | | | |
| 23/11/2020 – 29/11/2020 | 45,122 (5.4%) | 7,327 (1.2%) | 37,795 (15.3%) |
| 30/11/2020 – 06/12/2020 | 46,205 (5.5%) | 13,143 (2.2%) | 33,062 (13.4%) |
| 07/12/2020 – 13/12/2020 | 65,523 (7.8%) | 30,817 (5.2%) | 34,706 (14.1%) |
| 14/12/2020 – 20/12/2020 | 82,854 (9.9%) | 52,214 (8.8%) | 30,640 (12.4%) |
| 21/12/2020 – 27/12/2020 | 93,442 (11.1%) | 66,222 (11.2%) | 27,220 (11.0%) |
| 28/12/2020 – 03/01/2021 | 135,516 (16.1%) | 103,311 (17.4%) | 32,205 (13.0%) |
| 04/01/2021 – 10/01/2021 | 132,201 (15.8%) | 107,269 (18.1%) | 24,932 (10.1%) |
| 11/01/2021 – 17/01/2021 | 103,930 (12.4%) | 89,329 (15.1%) | 14,601 (5.9%) |
| 18/01/2021 – 24/01/2021 | 76,521 (9.1%) | 68,854 (11.6%) | 7,667 (3.1%) |
| 25/01/2021 – 31/01/2021 | 57,964 (6.9%) | 53,923 (9.1%) | 4,041 (1.6%) |

## 3.2 Hospitalisation

There were 27,710 hospitalisations within 1-14 days among the 592,409 SGTF patients (4.7%) and 8,523 among the 246,869 non-SGTF patients (3.5%). Only 911 (0.15%) SGTF and 399 (0.16%) non-SGTF patients died within 14 days without prior hospitalisation; hence, how deaths were treated would make little difference to the hospitalisation HR estimates.

*3.2.1 Hazard ratios*

The crude HR of hospitalisation within 1-14 days was 1.36 (95% CI 1.33 to 1.40) for SGTF compared to non-SGTF patients. Based on the base model, the HR of hospitalisation within 1-14 days was 1.52 (95% CI 1.47 to 1.57). The proportional hazards assumption was violated for this model (P<0.001). However, this may have reflected a high power to detect minor deviations from proportionality due to the large sample size, and the corresponding log-log plots showed approximately parallel curves (Appendix Figure A.1).

35,769 (98.7%) of the hospitalised patients were included in the analysis; the remaining 464 (1.3%) hospitalised patients were in single-individual strata and therefore uninformative. 183,491 (22.8%) of the non-hospitalised patients were uninformative due to being in a stratum where no individuals were hospitalised. Hence, a total of 655,323 patients (78.1%) were informative for the base model. Removing variables from the stratification set allowed the use of more observations and gave similar results, with HRs ranging from 1.40 to 1.51 (Appendix Table A.1).

Models including interactions between covariates and SGTF indicated no effect modification by sex (P=0.64), ethnicity (P=0.43), IMD (P=0.69), region of residence (P=0.21), or week (P=0.76). There was evidence that the effect was modified by age (P<0.001), with little difference in hospitalisation by SGTF status in those under 20 but rising to HRs in the range 1.45 to 1.65 in those aged 30 and older (Table 2; Appendix Table A.2).

When refitting the base model using a finer geographic stratification (LTLA) for the areas of residence instead of PHEC, 29,481 (81.4%) of the hospitalisations and 204,469 (24.4%) of the observations could be used. The adjusted HR was 1.52 (95% CI 1.47 to 1.58).

Extending the follow-up time, 46,371 SGTF (7.8%) and 16,654 non-SGTF (6.7%) patients had been hospitalised within 1-60 days of the first positive test. The crude HR of hospitalisation within 1-60 days was 1.17 (95% CI 1.15-1.19), and the corresponding fully stratified HR was 1.25 (95% CI 1.22 to 1.28). The proportional hazards assumption was violated (P<0.001). Consistently, the HR for SGTF varied with time since first positive test when allowing for a time-varying effect (P<0.001). The estimated HRs were 1.46 (95% CI 1.40 to 1.52) in days 1-7 after specimen, 1.62 (95% CI 1.54 to 1.70) in days 8-14, but subsequently close to 1.0 (range 0.91 to 1.03; Appendix Table A.3).

*3.2.2 Absolute risks*

Table 2 shows the HR estimates for SGTF from the absolute risk model, which were similar to those from the fully stratified base model, both overall and by age group. Based on this model, Table 2 and Figure 1 show estimates of the age-group-specific absolute risks of hospitalisation within 14 days after first positive test by SGTF status, at average levels of the potential confounders over all patients. The overall estimated absolute risk of hospitalisation after 14 days was 4.7% (95% CI 4.6% to 4.7%) for SGTF and 3.5% (3.4% to 3.5%) for non-SGTF patients. Most of the hospitalisations were in the first 14 days; Appendix Figure A.2 shows the corresponding graph to 60 days, which show a high hospitalisation rate over the first 14 days and an approximately constant low rate subsequently.

**Table 2** Hazard ratios of hospitalisation within 1-14 days for SGTF compared to non-SGTF patients.

| Model | Age group | SGTF status | n hospitalised/N (%) | Stratified base model* | Absolute risk model† | |
|---|---|---|---|---|---|---|
| | | | | HR (95% CI), SGTF vs non-SGTF | HR (95% CI), SGTF vs non-SGTF | Absolute 14-day hospitalisation risk (95% CI) |
| Base | Overall | SGTF | 27710/592409 (4.7%) | 1.52 (1.47 to 1.57) | 1.51 (1.47 to 1.55) | 4.7% (4.6% to 4.7%) |
| | | Non-SGTF | 8523/246869 (3.5%) | 1.00 (reference) | 1.00 (reference) | 3.5% (3.4% to 3.5%) |
| Age group-specific | <10 | SGTF | 288/31935 (0.9%) | 0.93 (0.70 to 1.25) | 0.97 (0.79 to 1.20) | 0.9% (0.8% to 1.0%) |
| | | Non-SGTF | 121/12281 (1.0%) | 1.00 (reference) | 1.00 (reference) | 1.0% (0.8% to 1.2%) |
| | 10-19 | SGTF | 472/63084 (0.7%) | 1.21 (0.99 to 1.49) | 1.18 (1.00 to 1.39) | 0.7% (0.7% to 0.8%) |
| | | Non-SGTF | 209/30646 (0.7%) | 1.00 (reference) | 1.00 (reference) | 0.7% (0.6% to 0.8%) |
| | 20-29 | SGTF | 2149/115296 (1.9%) | 1.29 (1.16 to 1.43) | 1.30 (1.19 to 1.42) | 1.9% (1.8% to 1.9%) |
| | | Non-SGTF | 707/45561 (1.6%) | 1.00 (reference) | 1.00 (reference) | 1.5% (1.4% to 1.7%) |
| | 30-39 | SGTF | 3964/118229 (3.4%) | 1.45 (1.34 to 1.58) | 1.41 (1.32 to 1.51) | 3.4% (3.3% to 3.5%) |
| | | Non-SGTF | 1216/47341 (2.6%) | 1.00 (reference) | 1.00 (reference) | 2.6% (2.4% to 2.7%) |
| | 40-49 | SGTF | 5162/102684 (5.0%) | 1.61 (1.50 to 1.74) | 1.59 (1.50 to 1.69) | 5.0% (4.9% to 5.2%) |
| | | Non-SGTF | 1429/41581 (3.4%) | 1.00 (reference) | 1.00 (reference) | 3.4% (3.3% to 3.6%) |
| | 50-59 | SGTF | 6734/93468 (7.2%) | 1.58 (1.48 to 1.69) | 1.58 (1.50 to 1.67) | 7.2% (7.0% to 7.4%) |
| | | Non-SGTF | 1902/38743 (4.9%) | 1.00 (reference) | 1.00 (reference) | 4.9% (4.7% to 5.1%) |
| | 60-69 | SGTF | 4733/44709 (10.6%) | 1.65 (1.53 to 1.79) | 1.63 (1.53 to 1.73) | 10.6% (10.3% to 10.9%) |
| | | Non-SGTF | 1361/19188 (7.1%) | 1.00 (reference) | 1.00 (reference) | 7.1% (6.8% to 7.5%) |
| | 70-79 | SGTF | 2653/15726 (16.9%) | 1.45 (1.32 to 1.60) | 1.49 (1.38 to 1.60) | 16.9% (16.3% to 17.5%) |
| | | Non-SGTF | 932/7477 (12.5%) | 1.00 (reference) | 1.00 (reference) | 12.5% (11.7% to 13.2%) |
| | 80+ | SGTF | 1555/7278 (21.4%) | 1.60 (1.41 to 1.82) | 1.50 (1.36 to 1.64) | 21.7% (20.7% to 22.7%) |
| | | Non-SGTF | 646/4051 (15.9%) | 1.00 (reference) | 1.00 (reference) | 16.2% (15.0% to 17.3%) |

\* The hazard ratios from the primary stratified models were estimated based on Cox regression stratified by the potential confounders 10-year age group, sex, ethnicity, IMD quintile, region of residence, and week of specimen; and using regression adjustment for the quantitative covariates age, IMD rank, and date of specimen.

† The secondary absolute risk model based on Cox regression stratified by SGTF status and age group only, and regressed on the remaining potential confounders. The 14-day absolute hospitalisation risks by SGTF status were estimated based on the absolute risk model, at the mean levels of the potential confounders.

**Figure 1.** Cumulative hospitalisation risk within 1-14 days after positive COVID-19 test, by age group. The risks were estimated based on a Cox regression model stratified by SGTF status and age group, adjusted for sex, IMD quintile, ethnicity, region of residence, and calendar week (potential confounders set to mean covariate levels).

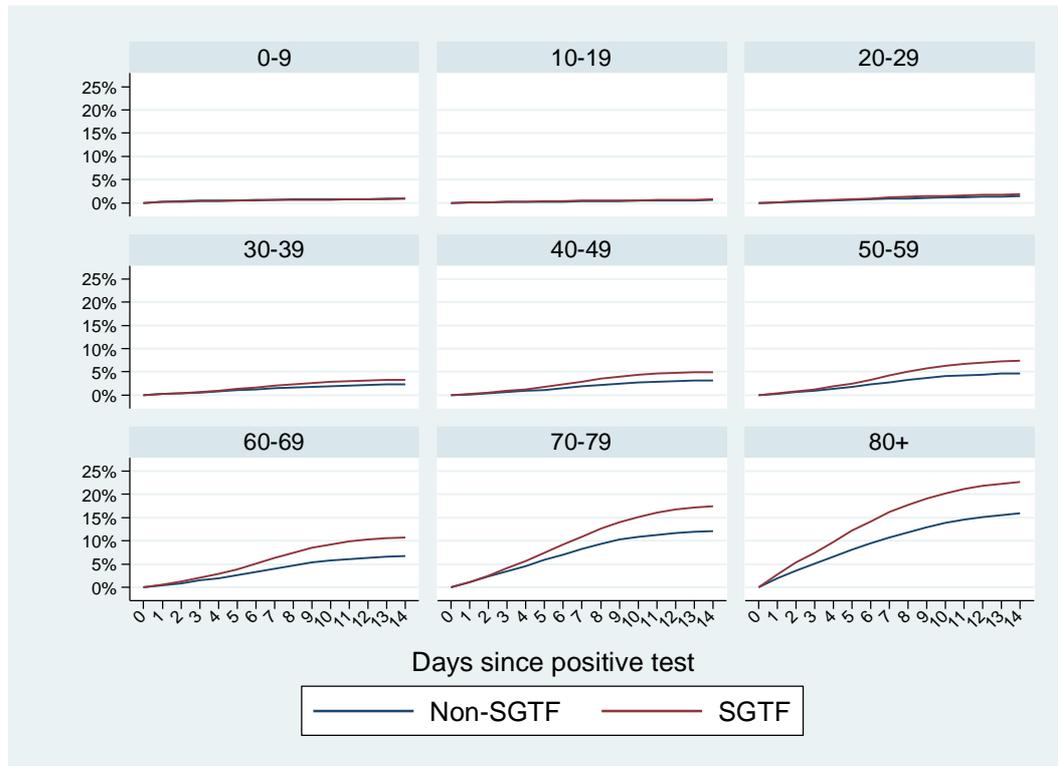

### 3.3 Mortality

There were 2,603 deaths within 28 days of positive test in the 592,409 SGTF patients (0.44%) and 899 deaths within 28 days in the 246,869 non-SGTF patients (0.36%). The crude HR of death was 1.22 (95% CI 1.13 to 1.31). After adjustment for all considered potential confounders the HR was 1.59 (95% CI 1.44 to 1.74).

## 4. Discussion

### 4.1 Principal findings

This retrospective analysis of COVID-19 patients identified through community testing in England indicated that the risk of hospitalisation within 14 days after a positive test was 1.52 (95% CI 1.47 to 1.57) times higher for diagnosed patients infected with the B.1.1.7 variant compared to patients infected with wildtype variants, after adjustment for age, sex, deprivation, ethnicity, region and week of diagnosis. Consistent with previously reported analyses of the study dataset,[6,7] the results indicated that B.1.1.7 is associated with a 1.59 (95% CI 1.44 to 1.74) times higher risk of death within 28 days.

The results indicated that the higher risk of hospitalisation may apply primarily to adults above the age of 30, and the risk was not found to be higher for SGTF compared to non-SGTF COVID-19 patients below the age of 20. This may reflect a similarly low risk of severe disease in younger and less comorbid individuals as previously reported for wildtype variants.[20]

We have provided absolute risk estimates of hospitalisation within 14 days by age group after adjustment for potential confounders. Under the model used to estimate the absolute risks, the B.1.1.7 variant can be estimated to have caused an excess 8941 hospitalisations within 14 days during the study period, based on a comparison of the observed hospitalisations with the expected hospitalisations under a counterfactual scenario in which patients with SGTF variants were hospitalised at the same rate as those with non-SGTF variants. However, this estimate ignores the excess transmissibility of B.1.1.7 [2,4,5] and hence likely underestimate the total number of hospitalisations attributable to B.1.1.7.

### 4.2 Strengths and limitations of the study

Strengths of this analysis include the use of the community-based dataset, which includes all COVID-19 patients identified through community testing in England. The large dataset allowed for the use of stratification to adjust for confounding due to individual-level demographic and socioeconomic factors, region and calendar period.

Although the observed increased risks of hospitalisation and mortality are both consistent with the hypothesis of increased severity for B.1.1.7 compared to wildtype SARS-CoV-2, hospitalisation may not be susceptible to the same confounding pathways as mortality. The calendar period in which the prevalence of the B.1.1.7 increased coincided with a general increase of COVID-19 diagnoses in the UK,[21] and initial outbreaks of B.1.1.7 were local to South East England and neighbouring regions.[5] It is possible that a high local pressure on the healthcare system might lead to higher mortality because of factors such as insufficient resources and staff per hospital-admitted patient. Hence, because the pressure on the healthcare system was higher in areas where B.1.1.7 was more prevalent, it is possible that those areas saw a somewhat higher mortality due to hospital over-burden. This could result in positive confounding, i.e. the mortality risk associated with B.1.1.7 may have been somewhat overestimated in this and previous analyses of community testing data in England.[6,7] In this and previous studies, hospital pressure was not directly adjusted for, and the potential confounding effect of local hospital pressure was only addressed indirectly through adjustment for time period and region.[6–8] In contrast, high local hospital pressure is not associated with the propensity that newly diagnosed COVID-19 patients experience disease sufficiently severe to require hospitalisation, and is unlikely to lead to a greater proportion of COVID-19 patients requiring hospital care being admitted to hospital. Hence, hospital pressure is unlikely to positively confound the association between SGTF status and hospital admission. Most likely, hospital pressure does not affect the propensity that a severely affected COVID-19 patient is admitted to hospital. If so, hospital pressure does not confound the association between SGTF status and hospitalisation. However, it is possible that due to severe local hospital over-burden, some severely affected COVID-19 patients who would otherwise have been hospitalised were not admitted, or admitted later than they otherwise would. Because of the time-and-region-specific overlap between high hospital pressure and high prevalence of the B.1.1.7 variant, this would however be expected to result in negative confounding and hence underestimation of the hospitalisation HR for B.1.1.7-infected patients. In the absence of full control for the potential confounding due to local hospital pressure in analyses of mortality risk, the association with hospitalisation corroborates the hypothesis that the B.1.1.7 variant is associated with more severe disease than wildtype variants.

Our study also has limitations. The study population included COVID-19 patients with known SGTF status from three laboratories providing community testing. These laboratories perform a large proportion of tests from across the country (41.0% of positive cases over the study period), but geographic distribution can vary depending on capacity from other available laboratories. These potential geographical differences in testing coverage were accounted for in the adjusted model. The community-based population excluded COVID-19 patients who were diagnosed after presentation directly to emergency or other healthcare services. Individuals who present directly to healthcare services may have more severe disease than those diagnosed through community testing. Community testing is largely self-selected, and we cannot control for the possibility that testing patterns may have differed between individuals infected with the B.1.1.7 variant and individuals infected with wildtype variants. Some evidence suggests that individuals infected with the B.1.1.7 variant are more likely to experience symptomatic disease compared to other infected individuals,[22] but it is unknown if the B.1.1.7 variant is also associated with more severe symptoms in the subset who experience symptoms.

Our analysis is limited by a lack of data on comorbidity and obesity, which are risk factors for hospitalisation with COVID-19.[20] Previous studies have however not noted any association between B.1.1.7 status and BMI [9] or comorbidity [8,9] in COVID-19 patients. Hence, we do not expect that these potential confounders were strongly associated with SGTF status. Furthermore, they were accounted for indirectly through age, sex, ethnicity and deprivation. In light of the finding that the hospitalisation risk for those infected with the B.1.1.7 variant increased with age, further research is however needed to understand whether the severity associated with the B.1.1.7 variant is modified by age, and associated factors such as comorbidity and obesity.

The analysis uses SGTF status which is a proxy test for the B.1.1.7 variant. However, any non-differential misclassification is likely to result in a small bias towards the null, and the available sequencing data indicate that the positive and negative predictive values of the SGTF test was >90% during the studied period.[2]

Healthcare recommendations for COVID-19 patients did not differ by SARS-CoV-2 variant, but the B.1.1.7 variant and the reports of its increased severity saw great media attention over the studied period and it is possible that this might have affected the healthcare seeking behaviour of COVID-19 patients. However, the assessment of SGTF was done for surveillance purposes and the SGTF status not routinely provided to patients or their healthcare providers, so we have no reason to believe that healthcare-seeking behaviour differed by SGTF status. Consistently, we observed no significant variation in the HR for SGTF compared to non-SGTF patients by calendar week.

Reporting delays might affect this analysis, particularly for the hospitalisation data from SUS which are not recorded until after the completion of a hospital episode. However, at least 108 days had elapsed between the patients' dates of positive test and the linkage with the hospitalisation data, which makes reporting delays likely to have had limited impact. Furthermore, the SGTF patients were on average diagnosed later in calendar time than the non-SGTF patients. Hence, any nondifferential underreporting of hospitalisations due to reporting delays may have resulted in underestimation of the hospitalisation risk predominantly in the SGTF patients, which might have resulted in a small underestimation of the true HR. Our analyses controlled for calendar week, which likely makes the effect of this limited. Additionally, the use of hospitalisation data from health service datasets that principally serve administrative purposes has the benefit of broad coverage.

The proportional hazards assumption of the Cox regression model was violated, but the corresponding log-log plots indicated that the deviation from proportionality was very minor within the first 14 days after the positive test. Consistently, when allowing for time-dependent effects the time-varying HRs in the first 14 days were similar to the overall HR of hospitalisation within 1-14 days from the primary analysis. By contrast, the time-dependent HRs of hospitalisation 15-60 days after the positive test from the additional analysis where we extended the follow-up time beyond 14 days were close to 1.0, and the hazards hence clearly non-proportional over this extended follow-up. These patterns likely reflect that the hospitalisation data were limited by a lack of information on the reason for admission and the analysis therefore based on hospitalisations due to any cause. Due to the narrow time interval considered in the primary analysis of hospitalisations between 1-14 days after a first positive SARS-CoV-2 test, most of these hospitalisations were likely due to COVID-19. Assuming that the background rate of hospitalisation due to non-COVID-19 causes were similar between SGTF and non-SGTF patients, the resulting non-differential misclassification of hospitalisations due to other causes may however have resulted in a slight underestimation of the cause-specific HR in days 1-14. Such misclassification is a likely explanation for the attenuation of the HR when we extended the follow-up period to 60 days, because non-COVID-19 causes may constitute a higher proportion of the reasons for the late hospitalisations.

**4.3 Comparison with other studies**

The estimated higher risk of hospitalisation is in line with a previous analysis that estimated a hazard ratio of hospitalisation of 1.34 (95% CI 1.07 to 1.66), based on follow-up of COVID-19 patients with sequencing-confirmed B.1.1.7 or wildtype SARS-CoV-2 in England.[13] That study observed only 120 hospitalisations in B.1.1.7-infected individuals which yielded a wide confidence interval. By contrast, in the present study we observed 27,710 hospitalisations in patients with SGTF which allowed us to provide higher precision estimates. The results are also consistent with a previous ecological analysis that estimated a hospitalisation relative risk of 1.7 based on hospitalisation patterns by the regional prevalence of SGTF variants in England,[11,12] and reported relative risks in the range 1.6—1.7 based on data from Scotland,[11] seven EU/EEA countries,[23] Denmark [24] and Canada.[10] This adds to the evidence of higher severity COVID-19 after infection with B.1.1.7 compared to wildtype SARS-CoV-2, as further indicated by its association with higher risk of intensive care admission [9,10,23] and death.[6–10]

Our finding of lower age-specific HR estimates in younger age groups are consistent with a previously reported age-specific adjusted odds ratio of hospitalisation of 1.0 for SGTF versus non-SGTF patients aged 0-19 in seven EU/EEA countries,[23] but contrasts with a study in Denmark which reported an adjusted odds ratio of 1.84 for SGTF patients aged 0-29.[24] Children and adolescents aged ≤18 who were hospitalised with COVID-19 in November 2020-January 2021 at King's College Hospital in London (where the local prevalence of B.1.1.7 was high) were reported to have had similar clinical severity and treatment requirements compared to those hospitalised in March-May 2020,[25] corroborating the suggestion that individuals in the youngest age groups experience no more severe disease if infected with B.1.1.7 compared to wildtype SARS-CoV-2.

**4.4 Conclusions**

The results from this large nationwide community testing cohort suggest that the risk of hospitalisation is higher for COVID-19 patients infected with the B.1.1.7 variant compared to wildtype variants, likely reflecting that the variant is associated with more severe disease. This higher severity may however be specific to adults above the age of 30, and further research is needed to identify if the severity is modified by factors associated with ageing. Taken together with the previous evidence of increased mortality and transmissibility, the results suggest that epidemics of the B.1.1.7 variant are likely to result in higher burden on the healthcare system in unvaccinated populations compared to epidemics of wildtype SARS-CoV-2.

**Acknowledgements:** We thank Alex Bhattacharya for help with the linkage of the datasets, Paula Blomquist for managing the SGTF data, Mary Sinnathamby for support with the statistical analysis, and Peter Kirwan for helpful suggestions during the manuscript preparation process.

**Contributors:**
TN, KAT, RJH, SRS, AC, DDA, GD and AMP designed the study.
KAT, RJH, JF, HA, AC and GD contributed to the linkage of the data sources.
RJH and KAT performed the statistical analysis.
TN, KAT, RJH, GD and AMP wrote the manuscript.
SRS, JF, HA, AC and DDA reviewed and revised the manuscript.
AMP, AC and DDA acquired funding.

**Competing interests:** GD's employer Public Health England has received funding from GlaxoSmithKline for a research project related to seasonal influenza and antiviral treatment. This project preceded and had no relation to COVID-19, and GD had no role in or received any funding from the project. All other authors declare no conflicts of interest.

**Funding:** This research was funded by the Medical Research Council (TN, DDA, AMP, Unit programme number MC_UU_00002/11, SRS Unit programme number MC_UU_00002/10); and via a grant from the MRC UKRI/DHSC NIHR COVID-19 rapid response call (TN, AC, DDA, AMP, grant ref: MC_PC_19074). This research was also supported by the NIHR Cambridge Biomedical Research Centre.

**Role of the funding sources:** The funders had no role in the study design; in the collection, analysis, and interpretation of data; in the writing of the report; and in the decision to submit the article for publication. All authors had full access to all of the data (including statistical reports and tables) in the study and take responsibility for the integrity of the data and the accuracy of the data analysis.

**References**

1    pangolin. grinch: global report investigating novel coronavirus haplotypes: B.1.1.7. 2021. https://cov-lineages.org/global_report_B.1.1.7.html (accessed 30 Mar 2021).
2    Public Health England. Investigation of novel SARS-COV-2 variant: Variant of Concern 202012/01. 2021. https://www.gov.uk/government/publications/investigation-of-novel-sars-cov-2-variant-variant-of-concern-20201201 (accessed 30 Mar 2021).
3    UK Government Prime Minister's Office. Prime Minister announces national lockdown. 2021. https://www.gov.uk/government/news/prime-minister-announces-national-lockdown (accessed 30 Mar 2021).
4    Volz E, Mishra S, Chand M, *et al.* Assessing transmissibility of SARS-CoV-2 lineage B.1.1.7 in England. *Nature* 2021;**593**:266–9. doi:10.1038/s41586-021-03470-x
5    Davies NG, Abbott S, Barnard RC, *et al.* Estimated transmissibility and impact of SARS-CoV-2 lineage B.1.1.7 in England. *Science (80- )* 2021;**372**:eabg3055. doi:10.1126/science.abg3055
6    Challen R, Brooks-Pollock E, Read JM, *et al.* Risk of mortality in patients infected with SARS-CoV-2 variant of concern 202012/1: matched cohort study. *BMJ* 2021;**372**:n579. doi:10.1136/bmj.n579
7    Davies NG, Jarvis CI, Edmunds WJ, *et al.* Increased mortality in community-tested cases of SARS-CoV-2 lineage B.1.1.7. *Nature* 2021;**593**:270–4. doi:10.1038/s41586-021-03426-1
8    Grint DJ, Wing K, Williamson E, *et al.* Case fatality risk of the SARS-CoV-2 variant of concern B.1.1.7 in England, 16 November to 5 February. *Eurosurveillance* 2021;**26**. doi:10.2807/1560-7917.ES.2021.26.11.2100256
9    Patone M, Thomas K, Hatch R, *et al.* Analysis of severe outcomes associated with the SARS-CoV-2 Variant of Concern 202012/01 in England using ICNARC Case Mix Programme and QResearch databases. *medRxiv* 2021.
10    Tuite AR, Fisman DN, Odutayo A, *et al.* COVID-19 Hospitalizations, ICU Admissions and Deaths Associated with the New Variants of Concern. Science Briefs of the Ontario COVID-19 Science Advisory Table. 2021;1(18). doi:10.47326/ocsat.2021.02.18.1.0
11    NERVTAG. NERVTAG: Update note on B.1.1.7 severity, 11 February 2021. 2021.https://assets.publishing.service.gov.uk/government/uploads/system/uploads/attachment_data/file/961042/S1095_NERVTAG_update_note_on_B.1.1.7_severity_20210211.pdf (accessed 30 Mar 2021).
12    Abbott S, Funk S. Population-level association between S-gene target failure and the relationship between cases, hospitalisations and deaths of Covid-19. 2021. https://github.com/epiforecasts/covid19.sgene.utla.rt/blob/main/severity-report.pdf
13    Dabrera G, Allen H, Zaidi A, *et al.* Assessment of Mortality and Hospital Admissions Associated with Confirmed Infection with SARS-CoV-2 Variant of Concern VOC-202012/01 (B.1.1.7) a Matched Cohort and Time-to-Event Analysis. *SSRN Electron J* Published Online First: 2021. doi:10.2139/ssrn.3802578
14    Clare T, Twohig KA, O'Connell A-M, *et al.* Timeliness and completeness of laboratory-based surveillance of COVID-19 cases in England. *Public Health* 2021;**194**:163–6. doi:10.1016/j.puhe.2021.03.012
15    UK Government Department of Health & Social Care. NHS Test and Trace statistics (England): methodology. 2021. https://www.gov.uk/government/publications/nhs-test-and-trace-statistics-england-methodology/nhs-test-and-trace-statistics-england-methodology (accessed 28 Apr 2021).
16    NHS Digital. Secondary Uses Service (SUS). 2021. https://digital.nhs.uk/services/secondary-uses-service-sus (accessed 30 Mar 2021).
17    NHS Digital. Emergency Care Data Set (ECDS). 2021. https://digital.nhs.uk/data-and-information/data-collections-and-data-sets/data-sets/emergency-care-data-set-ecds (accessed 30 Mar 2021).

18    Bhattacharya A, Collin SM, Stimson J, *et al.* Healthcare-associated COVID-19 in England: a national data linkage study. *medRxiv* Published Online First: 2021. doi:10.1101/2021.02.16.21251625
19    Brown AE, Heinsbroek E, Kall MM, *et al.* Epidemiology of Confirmed COVID-19 Deaths in Adults, England, March–December 2020. *Emerg Infect Dis* 2021;**27**:1468–71. doi:10.3201/eid2705.203524
20    Khawaja AP, Warwick AN, Hysi PG, *et al.* Associations with covid-19 hospitalisation amongst 406,793 adults: the UK Biobank prospective cohort study. *medRxiv* Published Online First: 2020. doi:10.1101/2020.05.06.20092957
21    Gov.uk. Coronavirus (COVID-19) in the UK. 2021. https://coronavirus.data.gov.uk/ (accessed 30 Mar 2021).
22    Office for National Statistics. Coronavirus (COVID-19) Infection Survey: characteristics of people testing positive for COVID-19 in England, 27 January 2021. 2021. https://www.ons.gov.uk/peoplepopulationandcommunity/healthandsocialcare/conditionsanddiseases/articles/coronaviruscovid19infectionsinthecommunityinengland/characteristicsofpeopletestingpositiveforcovid19inengland27january2021#symptoms-profile-by-cases-com (accessed 30 Mar 2021).
23    Funk T, Pharris A, Spiteri G, *et al.* Characteristics of SARS-CoV-2 variants of concern B.1.1.7, B.1.351 or P.1: data from seven EU/EEA countries, weeks 38/2020 to 10/2021. *Eurosurveillance* 2021;**26**. doi:10.2807/1560-7917.ES.2021.26.16.2100348
24    Bager P, Wohlfahrt J, Fonager J, *et al.* Increased Risk of Hospitalisation Associated with Infection with SARS-CoV-2 Lineage B.1.1.7 in Denmark. *SSRN Electron J* Published Online First: 2021. doi:10.2139/ssrn.3792894
25    Brookman S, Cook J, Zucherman M, *et al.* Effect of the new SARS-CoV-2 variant B.1.1.7 on children and young people. *Lancet Child Adolesc Heal* 2021;**5**:e9–10. doi:10.1016/S2352-4642(21)00030-4

Tommy Nyberg, *research associate* [a]*
Katherine A. Twohig, *senior epidemiology scientist* [b]
Ross J. Harris, *senior statistician* [c]
Shaun R. Seaman, *senior research associate* [a]
Joe Flannagan, *senior epidemiology scientist* [b]
Hester Allen, *principal epidemiology scientist* [b]
Andre Charlett, *head of department* [c]
Daniela De Angelis, *professor of statistical science for health* [a,c]
Gavin Dabrera, *consultant in public health medicine* [b]
Anne M. Presanis, *senior investigator statistician* [a]

[a] MRC Biostatistics Unit, University of Cambridge, Cambridge, United Kingdom.
[b] COVID-19 National Epidemiology Cell, Public Health England, London, United Kingdom.
[c] National Infection Service, Public Health England, London, United Kingdom.

* Corresponding author.

**Appendix Figure A.1.** Plots of the natural logarithm of time since positive test versus the -log-log transformed Kaplan-Meier estimate of the survival function. If the proportional hazards assumption made by the Cox regression model is true, these lines are expected to be parallel.

(a) By SGTF status.

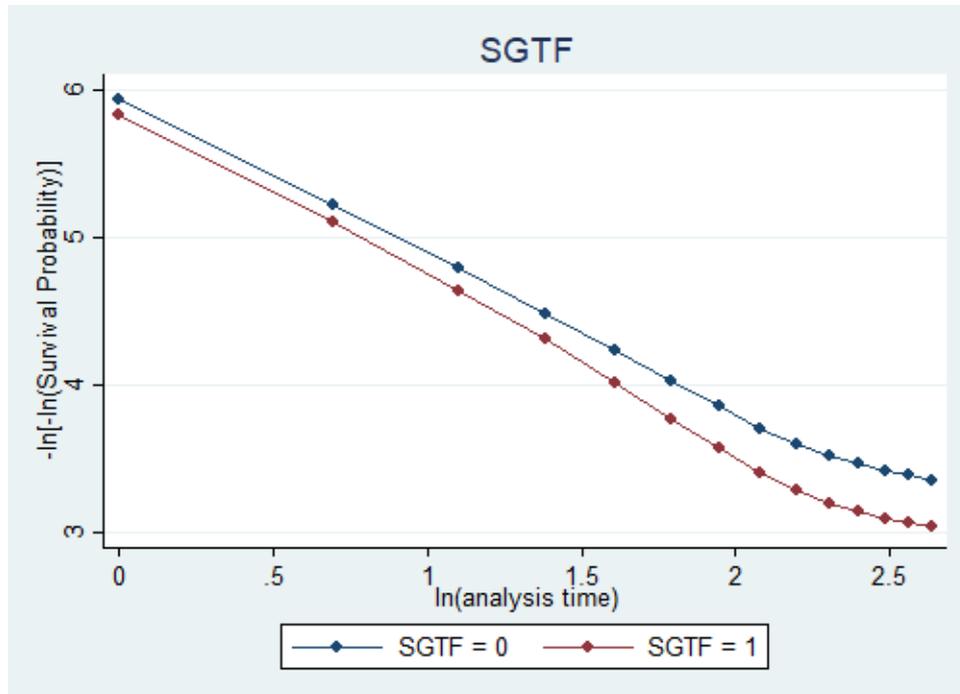

(b) By age group.

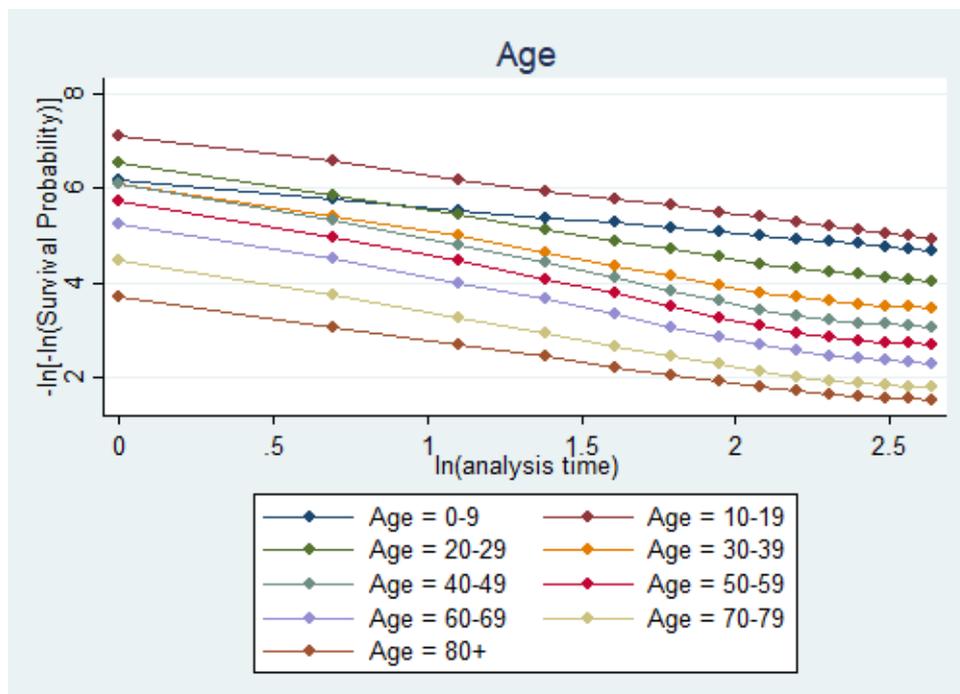

(c) By sex.

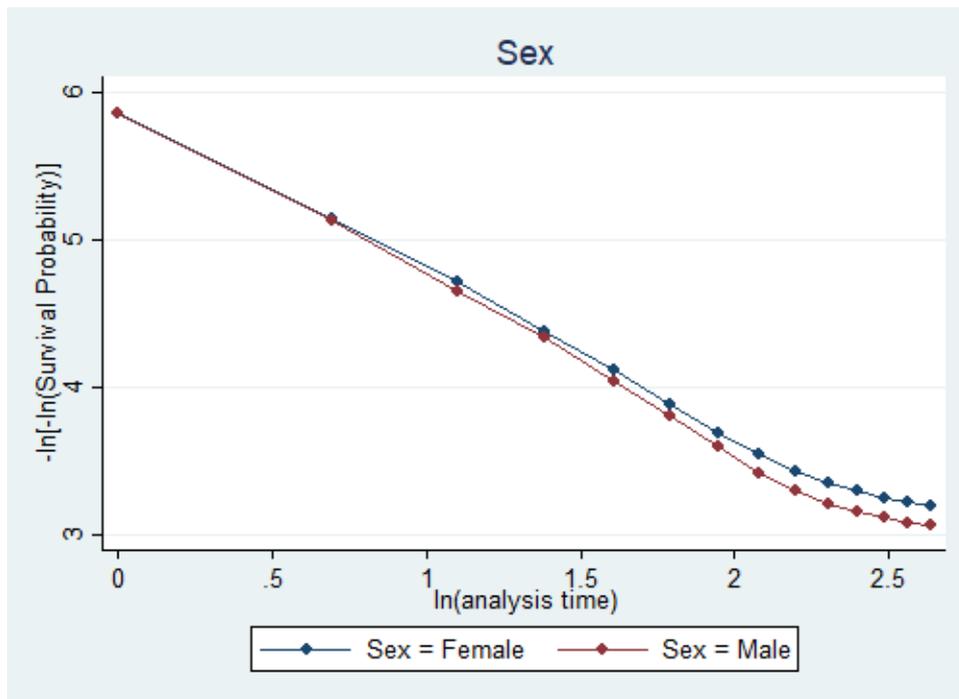

(d) By region of residence (PHEC).

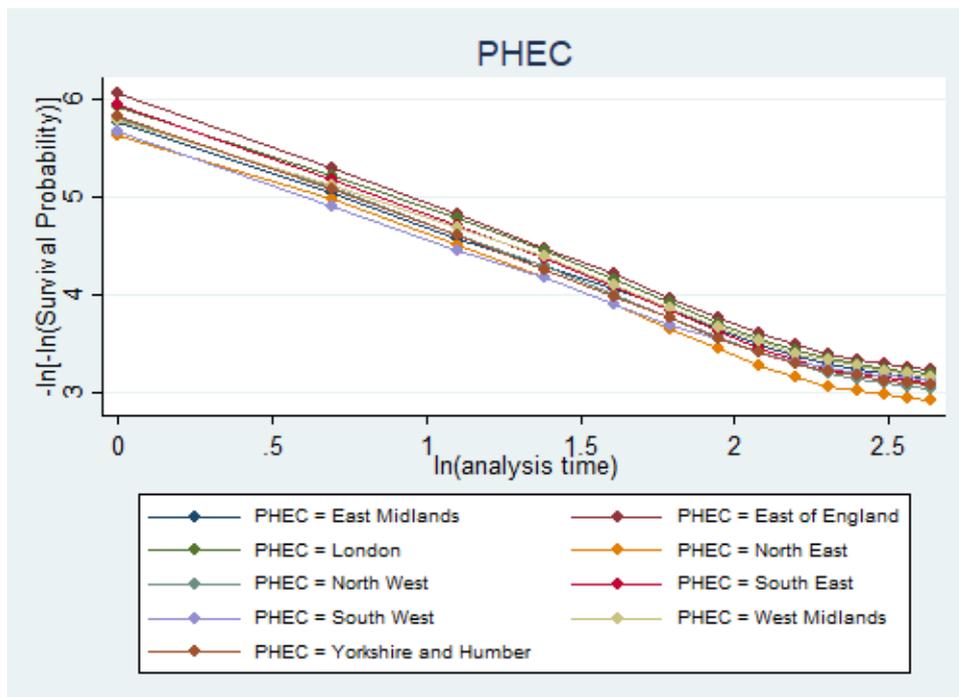

(e) By ethnicity.

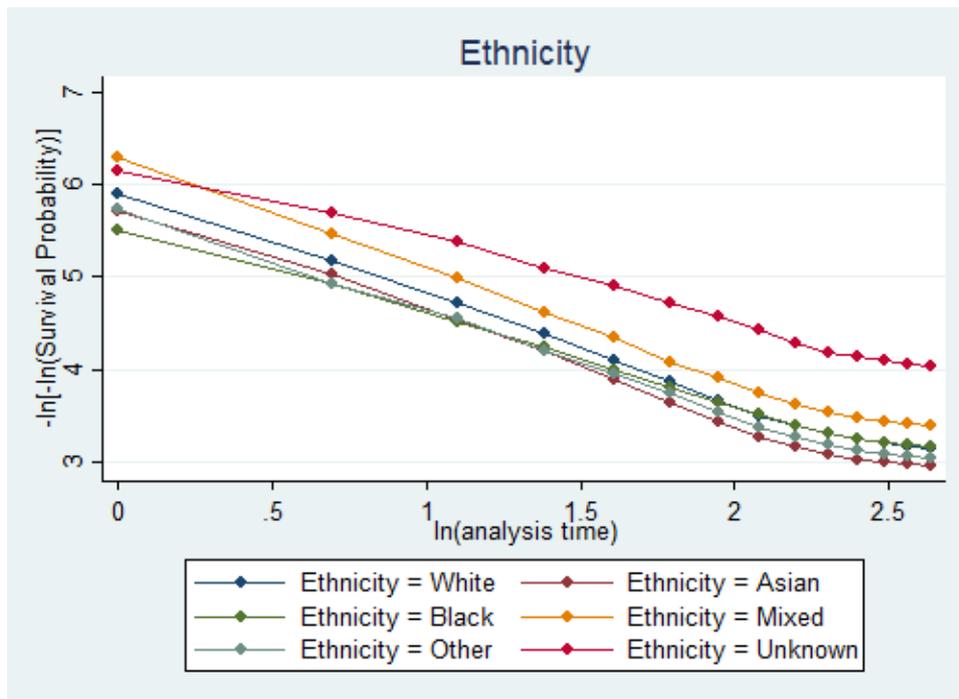

(f) By index of multiple deprivation quintile.

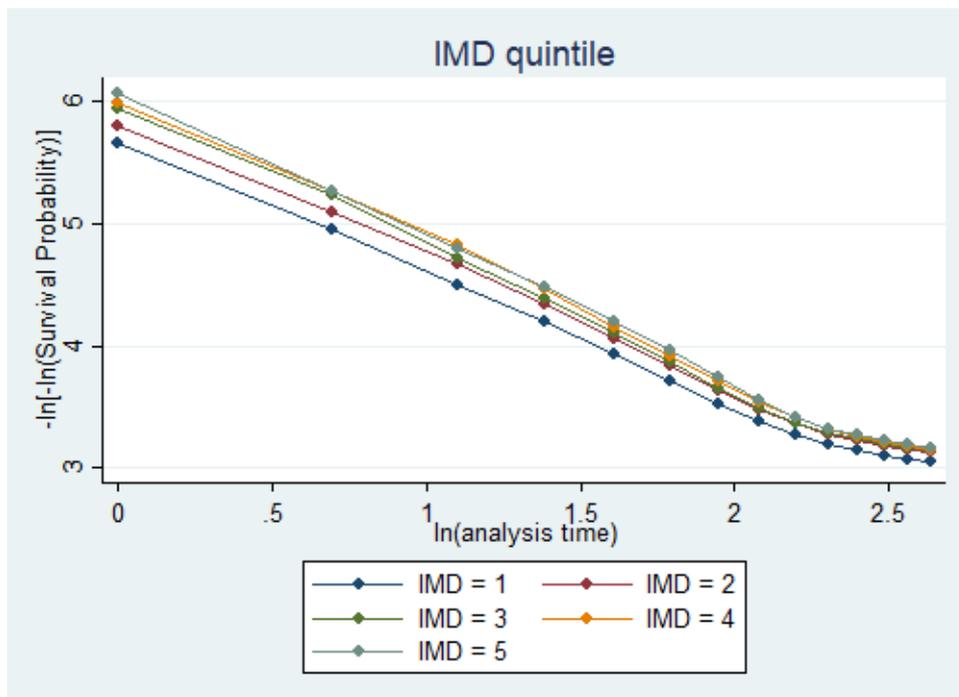

(g) By week of positive test.

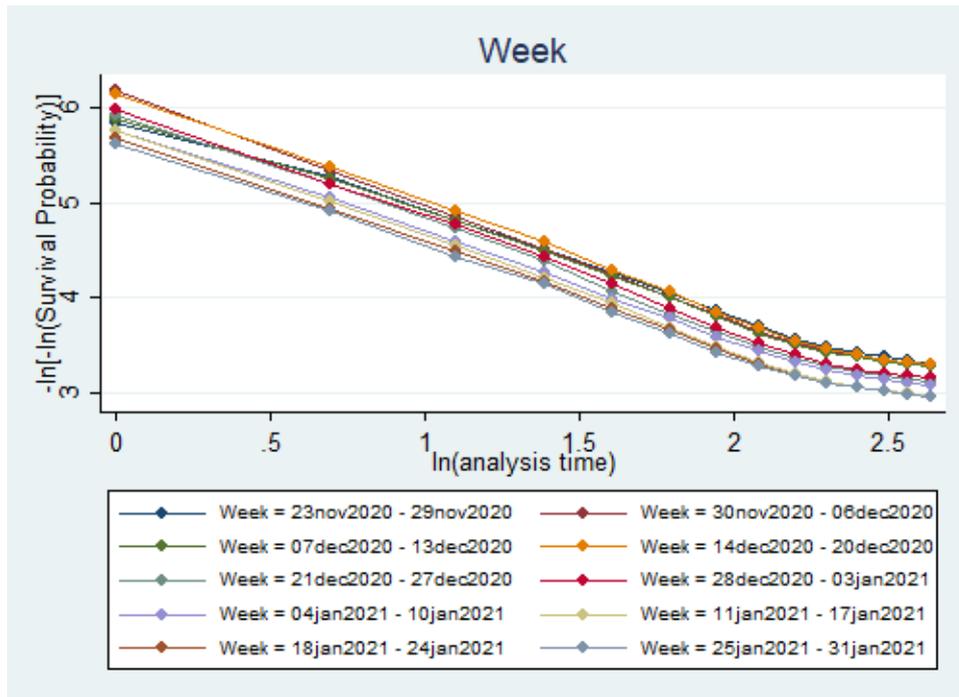

**Appendix Figure A.2.** Cumulative hospitalisation risk within 1-60 days after positive COVID-19 test based on a Cox regression model stratified by SGTF status and age group, adjusted for sex, IMD quintile, ethnicity, region of residence, and calendar week (adjustment covariates set to mean covariate levels).

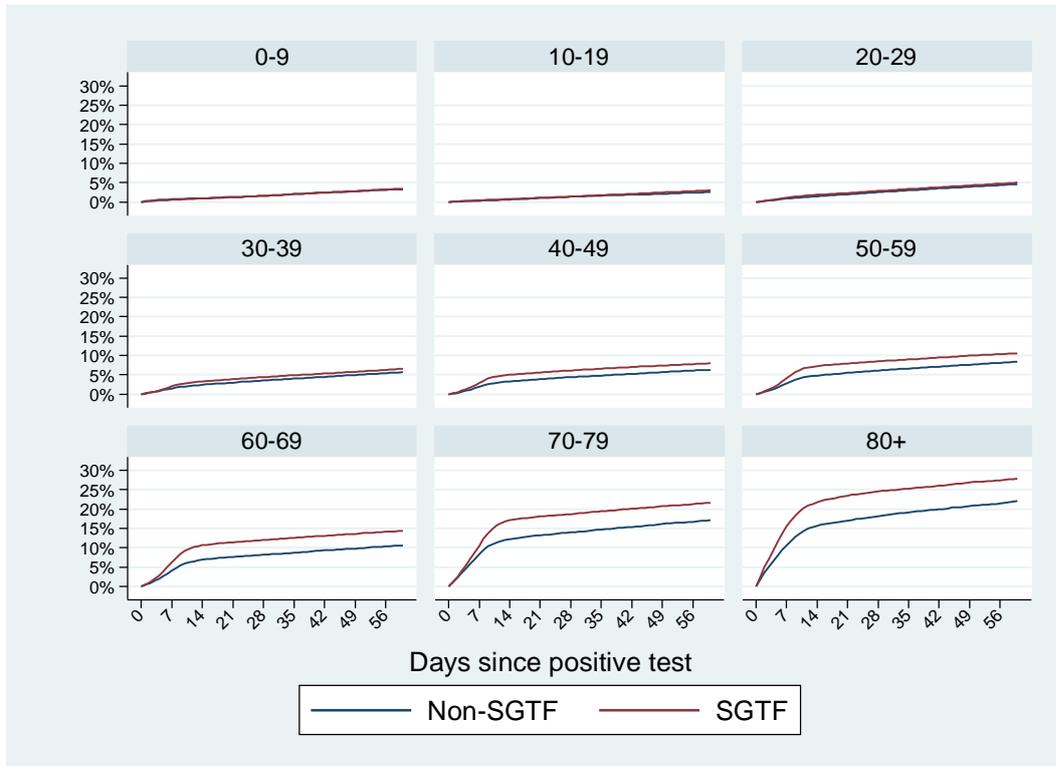

**Appendix Table A.1.** Number of hospitalisations and number of observations that were informative, and the corresponding hazard ratios of hospitalisation within 1-14 days for SGTF compared to non-SGTF patients, for the fully stratified "base model", and for corresponding reduced models with each stratification covariate excluded one-by-one. The hazard ratios were estimated stratified by the adjustment covariates 10-year age group, sex, ethnicity, IMD quintile, region of residence, and/or week of specimen; and using regression adjustment for the quantitative covariates age, IMD rank, and/or date of specimen.

| Stratification variables | N hospitalisations included (%) | N observations included (%) | SGTF HR (95% CI) |
|---|---|---|---|
| Stratification for all potential confounders: age, sex, ethnicity, IMD, PHEC, week ("base model") | 35769 (98.7%) | 655323 (78.1%) | 1.52 (1.47 to 1.57) |
| Excluding: age; Stratification for: sex, ethnicity, IMD, PHEC, week | 36226 (100.0%) | 812466 (96.8%) | 1.40 (1.35 to 1.44) |
| Excluding: sex; Stratification for: age, ethnicity, IMD, PHEC, week | 35991 (99.3%) | 720843 (85.9%) | 1.51 (1.46 to 1.55) |
| Excluding: PHEC; Stratification for: age, sex, ethnicity, IMD, week | 36198 (99.9%) | 791640 (94.3%) | 1.43 (1.39 to 1.47) |
| Excluding: ethnicity; Stratification for: age, sex, IMD, PHEC, week | 36221 (100.0%) | 750350 (89.4%) | 1.49 (1.44 to 1.53) |
| Excluding: IMD; Stratification for: age, sex, ethnicity, PHEC, week | 36137 (99.7%) | 772225 (92.0%) | 1.50 (1.45 to 1.54) |
| Excluding: week; Stratification for: age, sex, ethnicity, IMD, PHEC | 36204 (99.9%) | 805192 (95.9%) | 1.46 (1.43 to 1.50) |

**Appendix Table A.2.** Tests for interaction and hazard ratios of hospitalisation within 1-14 days for SGTF compared to non-SGTF patients by all considered covariates, based on stratified Cox regression. The hazard ratios were estimated stratified by the adjustment covariates 10-year age group, sex, ethnicity, IMD quintile, region of residence, and week of specimen; and using regression adjustment for the quantitative covariates age, IMD rank, and date of specimen.

| Covariate | Category | HR (95% CI) |
|---|---|---|
| Age group (test for interaction, P<0.001) | <10 | 0.93 (0.70 to 1.25) |
| | 10-19 | 1.21 (0.99 to 1.49) |
| | 20-29 | 1.29 (1.16 to 1.43) |
| | 30-39 | 1.45 (1.34 to 1.58) |
| | 40-49 | 1.61 (1.50 to 1.74) |
| | 50-59 | 1.58 (1.48 to 1.69) |
| | 60-69 | 1.65 (1.53 to 1.79) |
| | 70-79 | 1.45 (1.32 to 1.60) |
| | 80+ | 1.60 (1.41 to 1.82) |
| Sex (test for interaction, P=0.64) | Female | 1.53 (1.47 to 1.60) |
| | Male | 1.51 (1.45 to 1.58) |
| Region of residence (PHEC) (test for interaction, P=0.21) | East Midlands | 1.57 (1.37 to 1.79) |
| | East of England | 1.38 (1.22 to 1.55) |
| | London | 1.50 (1.38 to 1.62) |
| | North East | 1.58 (1.43 to 1.76) |
| | North West | 1.53 (1.44 to 1.63) |
| | South East | 1.61 (1.46 to 1.76) |
| | South West | 1.78 (1.50 to 2.12) |
| | West Midlands | 1.53 (1.41 to 1.65) |
| | Yorkshire & Humber | 1.41 (1.28 to 1.55) |
| Ethnicity (test for interaction, P=0.43) | White | 1.53 (1.48 to 1.59) |
| | Asian | 1.55 (1.43 to 1.67) |
| | Black | 1.39 (1.17 to 1.66) |
| | Mixed | 1.53 (1.13 to 2.08) |
| | Other | 1.39 (1.17 to 1.66) |
| | Unknown | 1.01 (0.62 to 1.64) |
| Index of multiple deprivation (test for interaction, P=0.69) | Quintile 1 – most deprived | 1.53 (1.44 to 1.62) |
| | Quintile 2 | 1.47 (1.38 to 1.57) |
| | Quintile 3 | 1.51 (1.41 to 1.62) |
| | Quintile 4 | 1.53 (1.42 to 1.65) |
| | Quintile 5 – least deprived | 1.59 (1.46 to 1.72) |
| Week of positive test (test for interaction, P=0.76) | 23/11/2020 – 29/11/2020 | 1.50 (1.29 to 1.74) |
| | 30/11/2020 – 06/12/2020 | 1.53 (1.34 to 1.74) |
| | 07/12/2020 – 13/12/2020 | 1.57 (1.42 to 1.75) |
| | 14/12/2020 – 20/12/2020 | 1.58 (1.44 to 1.74) |
| | 21/12/2020 – 27/12/2020 | 1.61 (1.48 to 1.76) |
| | 28/12/2020 – 03/01/2021 | 1.46 (1.36 to 1.57) |
| | 04/01/2021 – 10/01/2021 | 1.50 (1.39 to 1.62) |
| | 11/01/2021 – 17/01/2021 | 1.45 (1.32 to 1.58) |

| | |
|---|---|
| 18/01/2021 – 24/01/2021 | 1.56 (1.38 to 1.77) |
| 25/01/2021 – 31/01/2021 | 1.54 (1.30 to 1.82) |

**Appendix Table A.3.** Hazard ratios of hospitalisation within 1-60 days for SGTF compared to non-SGTF patients, based on stratified Cox regression allowing for time-varying (piecewise constant) HRs for SGTF status. The hazard ratios were estimated stratified by the adjustment covariates 10-year age group, sex, ethnicity, IMD quintile, region of residence, and week of specimen; and using regression adjustment for the quantitative covariates age, IMD rank, and date of specimen.

| Days after positive test | HR (95% CI) |
| --- | --- |
| 1-7 days | 1.46 (1.40 to 1.52) |
| 8-14 days | 1.62 (1.54 to 1.70) |
| 15-21 days | 1.03 (0.96 to 1.11) |
| 22-28 days | 0.91 (0.84 to 0.99) |
| 29-35 days | 0.99 (0.91 to 1.08) |
| 36-42 days | 0.97 (0.89 to 1.05) |
| 43-49 days | 1.02 (0.94 to 1.12) |
| 50-60 days | 0.95 (0.89 to 1.02) |